\documentclass[aps,prd,showpacs,superscriptaddress,groupedaddress]{revtex4}  
\usepackage{graphicx}  
\usepackage{dcolumn}   
\usepackage{bm}        
\usepackage{amssymb}   
\usepackage{commath}
\usepackage{color}

\begin{document}

\title{Cosmological Perturbations and Quasi-Static Assumption in $f(R)$ Theories}
\author{Mu-Chen Chiu}
\email{chiumuchen@gmail.com}
\affiliation{Scottish University Physics Alliance, Institute for
Astronomy, the Royal Observatory, University of Edinburgh, Blackford Hill, Edinburgh EH9 3HJ, UK}
 \affiliation{Shanghai Key Lab for Astrophysics, Shanghai Normal University, Shanghai 200234, China}

\author{Andy Taylor}
\affiliation{Scottish University Physics Alliance, Institute for
Astronomy, the Royal Observatory, University of Edinburgh, Blackford Hill, Edinburgh EH9 3HJ, UK}
\email{ant@roe.ac.uk}

\author{Chenggang Shu}
\affiliation{Shanghai Key Lab for Astrophysics, Shanghai Normal University, Shanghai 200234, China}\email{cgshu@shao.ac.cn}

\author{Hong Tu}
\affiliation{Shanghai Key Lab for Astrophysics, Shanghai Normal University, Shanghai 200234, China}\email{tuhong@shnu.edu.cn}        
\date{\today}

\begin{abstract}
 $f(R)$ gravity is one of the simplest theories of modified gravity to explain the accelerated cosmic expansion. Although it is usually assumed that the quasi-Newtonian approach (a combination of the quasi-static approximation and sub-Hubble limit) for cosmic perturbations is good enough to describe the evolution of large scale structure in $f(R)$ models, some studies have suggested that this method is not valid for all $f(R)$ models. Here, we show that in the matter-dominated era, the pressure and shear equations alone, which can be recast into four first-order equations to solve for cosmological perturbations exactly, are sufficient to solve for the Newtonian potential, $\Psi$, and the curvature potential, $\Phi$. Based on these two equations, we are able to clarify how the exact linear perturbations fit into different limits. We find that the Compton length controls the quasi-static behaviours in $f(R)$ gravity. In addition, regardless the validity of quasi-static approximation, a strong version of the sub-Hubble limit alone is sufficient to reduce the exact linear perturbations in any viable $f(R)$ gravity to second order. Our findings disagree with some previous studies where we find little difference between our exact and quasi-Newtonian solutions even up to $k=10 c^{-1} \mathcal{H}_0$.
\end{abstract}


\maketitle


\section{\label{sec:intro}Introduction}

Among all the possibilities to explain the observed accelerated expansion of the Universe with a modified
theory of Einstein's General Relativity (GR), $f(R)$
gravity (also dubbed fourth-order gravity) is
the simplest one~\cite{carroll04,hs07,Capozziello08,Linder09,cfps12,Sotiriou_n_Faraoni10,dFeTs05,Nojiri07}. In this class of theories, a new
function of the $Ricci$ scalar, $f(R)$, is included in the Einstein-Hilbert action. In this way, $f(R)$
models form a class of higher derivative gravity theories. This
is a natural extension of Einstein's General Relativity because
there is no prior reason to exclude these higher order terms from
the Lagrangian density~\cite{cfps12,Sotiriou_n_Faraoni10,dFeTs05}. Indeed, the higher order
terms of the $Ricci$ scalar, $R$, does appear in low energy effective
Lagrangians in string theory and other candidate theories of quantum
gravity~\cite{nojiri03}.

Phenomenally speaking, there are viable
$f(R)$ models that can not only yield a consistent and realistic
cosmology~\cite{nojiri06,song07b}, but also pass the solar
system test~\cite{hs07}. However, although some $f(R)$ models can produce a background evolution that is identical to the standard cosmology ($\Lambda$CDM), this same set of $f(R)$ models will differ from $\Lambda$CDM with regard to other phenomena, such as weak lensing, cluster abundance, cosmic microwave background (CMB), and baryon acoustic oscillations (BAO)~\cite{Schmidt09,yangetal10,LSSH12,planck15}. To understand these phenomena on large scales, cosmological perturbation theory is a significant building block. According to this theory, the original fluctuations were amplified beyond the scale of the horizon at the the end of inflationary epoch, and only after the horizon has grown to the size of the fluctuations, various structures we recognize today start to grow (see e.g.~\cite{Durrer93}). Moreover, we also learn from cosmological perturbation theory that there are three types of perturbation: tensor, vector, and scalar modes~\cite{kodama84}. Of these, the tensor perturbation will cause gravitational wave, and the vector perturbation will generate vorticity that decays with time and becomes entirely negligible.  This leaves the scalar perturbation as the only source that could contribute to the growth of cosmic structures.

Many papers have been devoted to develop the theory of cosmological linear perturbations within the framework of $f(R)$ gravity~\cite{zhang06,tsuji07,song07,bean07,libarrow07,pog08,cruz08,amen08,tsuji08,tsujietal08,ceo10,mat13}.
However, many of them simply formulate the linear perturbations in $f(R)$ models by taking the quasi-static assumption for granted~\cite{zhang06,tsuji07,amen08,tsuji08,tsujietal08}.
Even if several works do provide us with ways to solve for linear perturbations exactly, it is not always clear under what circumstance the exact linear perturbations could be approximated by the quasi-static solutions well~\cite{song07,bean07,libarrow07,pog08,ceo10}. Indeed, some studies suggest that the quasi-static approximation will break down outside the sound horizon of modified gravity~\cite{sb15}, or even, for certain $f(R)$ models, on sub-Hubble scales~\cite{cruz08}; others conclude that the quasi-static approximation is valid for the most practical $f(R)$ models either on the nonlinear~\cite{bhb15}, or near-Hubble scales~\cite{hpst12}.
To clarify the reason behind these seemingly contradictory results,~\cite{nbf14} analyzes the quasi-static approximation in $f(R)$ models within the Einstein frame, and find that the fast/slow-rolling behaviours of the cosmic background will determine the validity of the quasi-static approximation.

In this paper, we clarify when the quasi-static approximation will break down within the framework of metric $f(R)$ gravity alone. In other words, unlike~\cite{nbf14}, our analysis is purely based on linear perturbations of $f(R)$ gravity in the Jordan frame, and will not invoke the ideas of a fast/slow-rolling background as the explanation. For this purpose, we firstly formulate the exact linear perturbations in a simple and clarifying way, and then investigate how these exact linear perturbations fit into different limits analytically as well as numerically.

  The structure of the paper is organized as below. In Sec.~\ref{formalism}, we briefly
outline the basic formalism of $f(R)$ gravity, and its application to the cosmic background evolution. We then show how to reduce the modified Einstein equations and equations of conservations for $f(R)$ models into the pressure and shear equations from first principles, and prove analytically why these two equations are sufficient to solve for the Newtonian potential and spatial curvature. In Sec.~\ref{limit} we describe how linear perturbations in $f(R)$ gravity fit into various limits, and discuss the role of the sub-Hubble and the quasi-Newtonian assumptions in taking the limits. In Sec.~\ref{numerical}, we apply the numerical method to solve for the Newtonian potentials and spatial curvatures exactly from the pressure and shear equations, and compare the solutions with those obtained in various limits. We also use numerical solutions from our newly derived exact equations to rebut the concerns over the quasi-static approximations.
Finally, we outline a brief conclusion in Sec.~\ref{conclusion}.


\section{Formalism of  $f(R)$ Gravity}\label{formalism}

\subsection{Background evolution in $f(R)$ theories}{\label{2image}}
   In $f(R)$ gravity, the $Ricci$ Scalar, $R$, in the Einstein-Hilbert action  is generalized to a function of $R$, and the action can be written as
        \begin{equation}
                  S=\int d^{4}x
                  \sqrt{-g}\left[\frac{R+f(R)}{16\pi G}+\mathcal{L}_{m}\right],
         \label{action}
        \end{equation} where $\mathcal{L}_{m}$ is the Lagrangian for matter. According to this action, the effect of $f(R)$ models can be understood as an
additive correction to Einstein's standard gravity (GR; we will use GR and standard gravity interchangeably), so it is easy to understand how $f(R)$ gravity deviates from $\Lambda$CDM. This action also implies that
there is a scale to distinguish the dominated terms between Einstein's gravity
and $f(R)$ theory. In this notion of $f(R)$ gravity, the modified field equations will be
        \begin{equation}
                  G_{\mu\nu} +f_{R} R_{\mu\nu} -\frac{1}{2} g_{\mu\nu}f - \nabla_{\mu}\nabla_{\nu}f_{R}
                  +g_{\mu\nu} \nabla^{\sigma}\nabla_{\sigma}{f_R} = 8 \pi G T_{\mu\nu},
                  \label{fieldEqr}
       \end{equation}
       where $f_{R}\equiv df(R)/dR$.

If we only consider a flat, isotropic, and homogeneous Universe, it is still legitimate to apply the Robertson-Walker (RW) metric to $f(R)$ cosmology. Under this metric, the trace of the $Ricci$ tensor alone will yield the equality,
        \begin{equation}
          R= 6 a^{-2}\mathcal{H}^2+6a^{-1}{\mathcal{H}'}= 6 a^{-2}\mathcal{H}^2+6a^{-2}\dot{\mathcal{H}},
          \label{RicciFLW}
        \end{equation}
which is purely geometric and independent of the choice of theory of gravity (hereafter, we define ``$\;'\;$'' for $d/da$ with $\mathcal{H}\equiv \dot{a}/a={{d \ln{a}}/{d\eta}}$, where a dot is denoted as $d/d\eta$ and $\eta$ is conformal time). This equality tells us that as long as the RW metric is assumed, there will be one simple
        relation between the Hubble rate, $\mathcal{H}(a)$, and the $Ricci$ scalar, $R(a)$.
        Eq.~(\ref{RicciFLW}) also tells us that the $Ricci$ scalar is spatially
 invariant in an isotropic and homogeneous Universe; thus, only the temporal components
 of $\nabla_{\mu}\nabla_{\nu}f_{R}$ and
 $g_{\mu\nu} \nabla^{\sigma}\nabla_{\sigma}{f_R}$ in Eq.~(\ref{fieldEqr}) will be nonzero.
Indeed, like the Friedmann equation in Einstein's standard theory of gravity, the $0-0$ component of Eq.~(\ref{fieldEqr}) alone is
  sufficient to yield a modified Friedmann equation.

 Accordingly, after combining all the temporal components in the field equations, Eq.~(\ref{fieldEqr}), we will obtain the modified Friedmann
equation in $f(R)$ theory,
       \begin{equation}
                 \mathcal{H}^2-af_{R}\mathcal{H}\mathcal{H}'+{1 \over
                 6}f a^2+\mathcal{H}^{2}f_{RR}aR'={8 \pi G \over 3}a^2 \rho.
                 \label{FrEqf}
        \end{equation}
From this modified Friedmann equation and Eq.~(\ref{RicciFLW}), the Hubble function can then be expressed as a function of the $Ricci$ scalar,
        \begin{equation}
                 \mathcal{H}^2={{8 \pi G a^2 \rho/ 3} - a^2(f-f_R R)/6 \over {1+f_R+f_{RR}aR'}  },
                 \label{HubbleRatefR}
        \end{equation} where $\rho=3 \mathcal{H}_0^2 \Omega_m/8\pi G a^{3}$ in the matter-dominated Universe, and
        $\mathcal{H}_0$ is equal to the current value of the Hubble rate in $\Lambda CDM$.
        For clarity,  unless specifically noted, we choose units to have $c=\mathcal{H}_0=1$ hereafter.

        In order to tell how the background of the Universe in $f(R)$ gravity evolves with time, we need to solve for the $Ricci$
        scalar (or Hubble rate) as a function of time (or scale factor) explicitly.
        For this purpose, we find the most intuitive way is to recast Eq.~(\ref{FrEqf}) into a second order equation for the
        Ricci scalar.  After differentiating  Eq.~(\ref{HubbleRatefR}) with respect to $a$, and combining it with   Eq.~(\ref{FrEqf})
        and Eq.~(\ref{RicciFLW}) to eliminate $ \mathcal{H} \mathcal{H}'$, we obtain a new nonlinear second order differential equation;
        \begin{equation}
                 R''+\left({f_{RRR} \over f_{RR}}R'+{a R\over 6 \mathcal{H}^2}-{2\over a}\right)R'
                 +\left({{1+f_R}\over {3 f_{RR} \mathcal{H}^2}}\right)R+ {{8 \pi G \rho} \over f_{RR}\mathcal{H}^2}
                     -{4(1+f_{R})\over a^2 f_{RR}}=0,
                 \label{BackGroundEv}
        \end{equation} where $\mathcal{H}^2$ is given by Eq.~(\ref{HubbleRatefR}).
        Although Eq.~(\ref{BackGroundEv}) is mathematically equivalent to other equations written down in previous work for
        the cosmic expansion in $f(R)$ theory
        (see e.g.~\cite{dFeTs05,hs07,song07}), Eq.~(\ref{BackGroundEv})
        has the advantage that it shows the evolution of the $Ricci$ scalar as an explicit solution of a second order equation.
        Like all second order differential equations, a negative coefficient of the second term in Eq.~(\ref{BackGroundEv}) will
        lead to a positive feedback system, which is highly unstable if the sign of this coefficient never changes.
        The role of this term becomes even clearer when the stability of $f(R)$ theory is considered~\cite{Chiu13}.


\subsection{\label{Newtonian Gauge}Cosmological perturbations in conformal Newtonian gauge}

                For any covariant linear perturbation around a flat, isotropic and homogeneous
Universe (see e.g.~\cite{kodama84,bardeen80}), we can decompose the perturbation into scalar, vector, and tensor modes, with a harmonic
expansion on a 3D sphere (for a 4D universe).  Because only scalar perturbations will contribute to the growth of structure, we do not consider vector and tensor perturbations in this paper.  In addition, we will only focus our analysis on the conformal Newtonian gauge, which will yield the Newtonian-like equations, and is popular in the literature of structure formation and weak lensing~\cite{ma95}.  For linear perturbations in a generic gauge in $f(R)$ gravity, we refer the readers to~\cite{bean07}.

Following the notation of~\cite{ma95}, we write down the RW metric in the Newtonian gauge as
        \begin{equation}
            ds^2=-a^2(1+2\Psi)d\eta^2+a^2(1-2\Phi)dx^2,
            \label{RW_pert}
        \end{equation} where $\Psi$ is called the Newtonian potential and $\Phi$ is the spatial curvature. We then apply this metric to the conservation equations, and obtain
        \begin{eqnarray}
            &&\dot{\delta}+(1+{\textit{w}})(\theta-3 \dot{\Phi})+3\mathcal{H}({c_s^2
            -{\textit{w}}})\delta=0,
            \label{dE-M01}\\
            &&\dot{\theta}+\mathcal{H}\left(1-3{\textit{w}}+{\dot{\textit{w}}\over 1+\textit{w}}\right)\theta
            -{k^2c_s^2 \over {1+\textit{w}}} \delta-k^2 \Psi=0.
            \label{dE-M02}
        \end{eqnarray} Here, $\delta$ is defined as the density perturbation, $\theta$ as the amplitude of spatial velocity, the speed of sound is $c_s^2 = \delta P/\delta \rho$, and $P/\rho=\textit{w}$ is the equation of state. Similarly, we can apply the metric, Eq.~(\ref{RW_pert}), to the perturbed modified Einstein equation. Accordingly, we will derive the Poisson equation,
        \begin{equation}
            (1+f_R)[-k^2(\Phi+\Psi)-3\mathcal{H}(\dot{\Phi}+\dot{\Psi})+(3\dot{\mathcal{H}}-6\mathcal{H}^2)\Psi
            -3\dot{\mathcal{H}}\Phi] +\dot{f_R}(-9\mathcal{H}\Psi+3\mathcal{H}\Phi-3\dot{\Phi})=8
            \pi G \rho a^2 \delta+(k^2-3\dot{\mathcal{H}})\tilde{\sigma}+3\mathcal{H}\dot{\tilde{\sigma}},
            \label{EinEq00Newf}
         \end{equation}
the pressure equation,
         \begin{eqnarray}
            (1+f_R)[\ddot{\Psi}&+&\ddot{\Phi}+3\mathcal{H}(\dot{\Psi}+\dot{\Phi})+3\dot{\mathcal{H}}\Psi
            +(\dot{\mathcal{H}}+2\mathcal{H}^2)\Phi]  \nonumber \\
            &+&\dot{f_R}(3\mathcal{H}\Psi-\mathcal{H}\Phi+3\dot{\Psi})+\ddot{f_R}(3\Psi-\Phi)={8\pi G a^2 \delta P}+ \Big(2\mathcal{H}^2
            +{\mathcal{H}}-{2\over 3} k^2\Big){\tilde{\sigma}}+\mathcal{{H}}\dot{\tilde{\sigma}}-\ddot{\tilde{\sigma}},
            \quad\quad\quad
              \label{EinEqiif}
         \end{eqnarray}
the momentum equation,
         \begin{equation}
            (1+f_R)[\dot{\Psi}+\dot{\Phi}+\mathcal{H}(\Psi+\Phi)]+ \dot{f_R}(2\Psi-\Phi)=8
            \pi G \rho a^2 \theta/k^2+\mathcal{H}\tilde{\sigma}-\dot{\tilde{\sigma}},
            \label{EinEq0if}
         \end{equation}
and the shear equation,
\begin{equation}
            (\Psi-\Phi)+{\tilde{\sigma}\over{1+f_R}}= -{2 f_{RR}\over a^2 (1+f_R)}[-6(\dot{\mathcal{H}}+\mathcal{H}^2)\Psi-3\mathcal{H}\dot{\Psi}
            +k^2\Psi-9\mathcal{H}\dot{\Phi}-3\ddot{\Phi}-2k^2\Phi],
            \label{EinEqijNewf}
\end{equation}
where $ \delta P$ is the perturbation of pressure density. We also define
\begin{equation}
\tilde{\sigma}=12\pi G a^2(\rho+P) \sigma/k^2,
\label{sigma}
\end{equation} where $\sigma$ is anisotropic stress. Unlike the their counterparts in GR, in $f(R)$ theories the anisotropic stress not only exists in the shear equation, but also appears in Poisson equation, the pressure equation, and the momentum equation. However, when $f(R)$ is close to zero, we can eliminate the anisotropic stress terms from the Poisson, the pressure, and the momentum equations by substituting $\tilde{\sigma}$ in the shear equation. This will make $\tilde{\sigma}$ only appear in the shear equation, and recover the GR regime.
It is also worth noting that in these equations,
\begin{eqnarray}
\dot{f_R}&=&f_{RR}\dot{R}, \\
\ddot{f_R}&=&f_{RRR}\dot{R}^2+f_{RR}\ddot{R},
\end{eqnarray}
where $f_{RR}=d^2f/dR^2$ and $f_{RRR}=d^3f/dR^3$; hence $f_R$, $f_{RR}$, and $f_{RRR}$ all affect the linear perturbations in $f(R)$ gravity in an explicit way.

Since so far we have only assumed a flat, isotropic and homogeneous Universe, the conservation equations, Eq.~(\ref{dE-M01})-(\ref{dE-M02}), and modified Einstein equation, Eq.~(\ref{EinEq00Newf})-(\ref{EinEqijNewf}), can also be applied to the radiation-dominated era as well as the matter-dominated era. Although it is quite straightforward to derive Eq.~(\ref{EinEq00Newf})-(\ref{EinEqijNewf}), they do not seem to have appeared in the literature previously.

\subsection{\label{Newtonian Gauge}Matter-dominated era}

Considering the success of the standard cosmological model ($\Lambda$CDM) on the cosmic microwave background (CMB)~\cite{Spergel07,planck14}, we follow~\cite{song07} to take the assumption that $f(R)$ models will recover GR before the CMB is formed. Accordingly, we will only analyze the linear perturbations in the matter-dominated era.

In the matter-dominated era, radiation density is ignored compared to matter density, so that $c_s^2 \equiv \delta P/\delta \rho \simeq P/\rho=\textit{w}=0$, and the conservation equations, Eq.~(\ref{dE-M01})-(\ref{dE-M02}), can be reduced into
      \begin{equation}
            \ddot{{\delta}}+\mathcal{H}\dot{\delta}+k^2\Psi-3\mathcal{H}\dot{\Phi}-3\ddot{\Phi}=0.
            \label{eq:cons}
        \end{equation}
In addition, it is also usually assumed that the Universe is free of anisotropic stress in the matter-dominated era, so it is legitimate to reduce the modified Einstein equations, Eq.~(\ref{EinEq00Newf})-(\ref{EinEqijNewf}), further by setting $\sigma=0$.
Indeed, if we take $\sigma=0$, the modified Einstein equations, Eq.~(\ref{EinEq00Newf})-(\ref{EinEqijNewf}), will be identical to the field equations derived by~\cite{cruz08} to solve for matter density perturbation. However, what differs from~\cite{cruz08} is that we do not attempt to recast all six equations (2 conservation equations and 4 fields equations) into one single differential equations of matter density contrast. As we are going to show, it will become a much easier task if we solve for $\Psi$ and $\Phi$ first, and then put these two potentials back into the Poisson equation and the momentum equation to obtain evolutions of $\delta$ and $\theta$, because the only equations required to solve for both potentials are the pressure equation,
         \begin{equation}
            (1+f_R)[\ddot{\Psi}+\ddot{\Phi}+3\mathcal{H}(\dot{\Psi}+\dot{\Phi})+3\dot{\mathcal{H}}\Psi
            +(\dot{\mathcal{H}}+2\mathcal{H}^2)\Phi] +\dot{f_R}(3\mathcal{H}\Psi-\mathcal{H}\Phi+3\dot{\Psi})+\ddot{f_R}(3\Psi-\Phi)=0,
              \label{EinEqii}
         \end{equation}
and the shear equation,
         \begin{equation}
            (\Psi-\Phi)=-{2 f_{RR}\over a^2 (1+f_R)}[-6(\dot{\mathcal{H}}+\mathcal{H}^2)\Psi-3\mathcal{H}\dot{\Psi}
            +k^2\Psi-9\mathcal{H}\dot{\Phi}-3\ddot{\Phi}-2k^2\Phi].
            \label{EinEqijNew}
         \end{equation}

To show the redundancy of the conservation and fields equations in $f(R)$ gravity, we firstly combine the conservation of momentum equation, Eq.~(\ref{dE-M02}), and the momentum equation, Eq.~(\ref{EinEq0if}), by eliminating $\theta$ in both equations. We will thus obtain
        \begin{equation}
            (1+f_R)[\ddot{\Psi}+\ddot{\Phi}+3\mathcal{H}(\dot{\Psi}+\dot{\Phi})+\dot{\mathcal{H}}(\Psi+\Phi)
            +2\mathcal{H}^2(\Psi+\Phi)]
            +\dot{f_R}(5\mathcal{H}\Psi-\mathcal{H}\Phi+3\dot{\Psi})+\ddot{f_R}(2\Psi-\Phi)=a^2 8\pi G \rho\Psi.
              \label{EinEqii_alt}
         \end{equation}
Then it is straightforward to prove that the difference between this equation and the pressure equation, Eq.~(\ref{EinEqii}), tells us no more than how the cosmic background evolves. In other words, in terms of the evolutions of $\Psi$ and $\Phi$, information given from the conservation equation for momentum and the momentum equation is equivalent to that derived from the pressure equation alone. Because of this equivalence, we call Eq.~(\ref{EinEqii_alt}) the ``alternative pressure equation".
Similarly, by combining the conservation equation for the density perturbation, Eq.~(\ref{dE-M01}), Poisson equation,
         Eq.~(\ref{EinEq00Newf}), and the shear equation, Eq.~(\ref{EinEqijNew}), we can cancel out $\delta$, and show that a combination of these three equations does not tell us
         more than Eq.~(\ref{EinEqijNew})~(the shear equation). Hence, we conclude that the pressure equation, Eq.~(\ref{EinEqii}), and the shear equation, Eq.~(\ref{EinEqijNew}), alone are sufficient to solve for $\Phi$ and $\Psi$. The ``alternative pressure equation", Eq.~(\ref{EinEqii_alt}), show explicitly  how matter density will affect evolutions of both potentials. On the contrary, if we choose the pressure equation, Eq.~(\ref{EinEqii}), over the ``alternative pressure equation", Eq.~(\ref{EinEqii_alt}), then the role of matter density will become implicit; that is, the evolutions of both potentials will be determined purely by the Hubble rate, $\mathcal{H}$, and the $Ricci$ scalar, $R$.

\section{Compton length and Effect of $f(R)$ gravity}\label{limit}

In the previous section, we have shown that the coupled pressure and shear equations are all we need in order to solve for $\Psi$ and $\Phi$ exactly. To understand how these equations fit into various limits of cosmological perturbations in $f(R)$ theories, we define a scale length,
         \begin{equation}
         \lambda_f\equiv\sqrt{f_{RR}\over a^{2}(1+f_R)},
         \label{fR_length}
         \end{equation}
which is easily  recognizable in the shear equation, Eq.~(\ref{EinEqijNew}). Based on this scale length, the elements of the shear equation can be classified into two groups: one is proportional to the scale dependent $1+k^2 \lambda_f^2$; the other is proportional to just $\lambda_f^2$, where $k^{-1}$ and $\lambda_f$ are in the unit, $\mathcal{H}_0/c$. It is worth noting that although derived from a different set of equations, $\lambda_f$ looks very similar to the Compton wavelength in~\cite{hs07}, which we compare with in Appendix~\ref{App_B}, or the lengthscale defined in~\cite{pog08}.

Before we discuss how these factors will affect linear perturbations in $f(R)$ gravity, we would like to clarify our definition of the so called~\textit{quasi-static approximation}, which is sometimes confusing in the existing literature (cf.~\cite{cruz08,mat13,hpst12,nbf14,bflm14}). According to our definition, the~\textit{quasi-Newtonian approximation} contain two parts:
 the sub-Hubble limit (~$k \gg \mathcal{H}$~) and the quasi-static approximation (~$ \dot{\abs{X}}\lesssim \mathcal{H}\abs{X} $~), where $X$ might be $\mathcal{H}$, $\Psi$ or $\Phi$. Not following the notations in~\cite{pog08,cruz08,mat13,hpst12,nbf14,bflm14}, we call $k \gg \mathcal{H}$ as the sub-Hubble limit rather than the sub-horizon limit because $\mathcal{H}$ is not a horizon but the Hubble length, as well as because we would like to emphasize the difference between this lengthscale and the particle horizon of modified gravity discussed in~\cite{sb15}.

 It shall be kept in mind that although $\mathcal{H}^2$ might become large at the early epoch, its effect must be balanced out by $\lambda_f^2$, otherwise $f(R)$ gravity will not match $\Lambda$CDM in the early Universe. By contrast, when the effects of $f(R)$ gravity start to kick-in at the more recent epoch, $\mathcal{H}$ should have become compatible with the order of $\mathcal{H}_0$. Based on this fact, we also introduce a weaker version of the sub-Hubble approximation, $k \gg \mathcal{H}_0$. Any $k$ that satisfies the sub-Hubble limit must also satisfy this weaker version of the sub-Hubble limit, but not vice versa.

\subsection{Reduced second-order differential equations}\label{qs:p}

Without the need to invoke the quasi-static approximation, when $1+k^2 \lambda_f^2 \gg k_X^2 \lambda_f^2$, we can reduce the shear equation, Eq.~(\ref{EinEqijNew}), into a simple relation between $\Psi$ and $\Phi$ by dropping terms proportional to $\lambda_f^2$. Here we introduce a wavenumber,
  \begin{equation}
    k_X^2=\rm{max}\{\mathcal{H}^2,\mathcal{H}\dot{X}/X,\ddot{X}/X\},
  \end{equation}
  which is defined by the maximum value of expansion rate of Universe or rate of change of $X$. Again, $X$ is defined as $\mathcal{H}$, $\Psi$ or $\Phi$; however, our discuss below will only focus on $X$ as representing $\Psi$ or $\Phi$, because for any $f(R)$ model that yields a similar cosmic acceleration to $\Lambda$CDM, $\mathcal{H}^2\sim\dot{\mathcal{H}}$$\sim\ddot{\mathcal{H}}$. In terms of the scale length, $\lambda_f$, Eq.~(\ref{EinEqijNew}) reduces to
         \begin{equation}
         (1+4 k^2{\lambda^2_f}){\Phi}=(1+2 k^2{\lambda^2_f}){\Psi}.
         \label{phi_psiRel}
         \end{equation}
Note that this immediately gives us the gravitational slip between the potentials, $\Phi/\Psi$.
This relation then can be put back into the alternative pressure equation, Eq~(\ref{EinEqii_alt}), to yield
the reduced second-order differential equations,
         \begin{eqnarray}
         {\ddot{\Psi}}&+&\left(3\mathcal{H}+A_1-A_3\right){\dot{\Psi}}
           +\left(\dot{\mathcal{H}}+2\mathcal{H}^2+A_2+B_{\psi}\right){\Psi}=0,
         \label{psi_2nd}
         \\
         {\ddot{\Phi}}&+&\left(3\mathcal{H}+A_1+A_3\right){\dot{\Psi}}
           +\left(\dot{\mathcal{H}}+2\mathcal{H}^2+A_2-B_{\phi}\right){\Psi}=0,
           \label{phi_2nd}
         \end{eqnarray}
         where we define
         \begin{eqnarray}
         A_1&=&{\frac{3\dot{f}_R(1+4k^2\lambda_f^2)}{(1+f_R)(2+6 k^2\lambda_f^2)}},
           \label{a1_c}
           \\
         A_2&=&{\frac{4\mathcal{H}\dot{f}_R+\ddot{f}_R}{(1+f_R)(2+6 k^2\lambda_f^2)}}+{\frac{k^2\lambda_f^2(9\mathcal{H}\dot{f}_R+3\ddot{f}_R)}{(1+f_R)(1+3 k^2\lambda_f^2)}}
         -{\frac{(1+4k^2\lambda_f^2)a^2 4\pi G\rho}{(1+3 k^2\lambda_f^2)(1+f_R)}},
         \label{a2_c}
           \\
         A_3&=&{\frac{4k^2{\lambda_f}\dot{\lambda}_f}{(1+2 k^2\lambda_f^2)(1+3 k^2\lambda_f^2)}},
          \label{a3_c}
         \end{eqnarray}
         \begin{eqnarray}
         B_{\psi}&=&{\frac{2k^2(4k^2{\lambda_f}^2\dot{\lambda}_f^2-3\mathcal{H}{\lambda}_f\dot{\lambda}_f
          -\dot{\lambda}_f^2-{\lambda}_f\ddot{\lambda}_f)}{(1+4 k^2\lambda_f^2)(1+3 k^2\lambda_f^2)}},
          \label{bpsi_c}
           \\
         B_{\phi}&=&{\frac{2k^2(8k^2{\lambda_f}^2\dot{\lambda}_f^2-3\mathcal{H}{\lambda}_f\dot{\lambda}_f
          -\dot{\lambda}_f^2-{\lambda}_f\ddot{\lambda}_f)}{(1+2 k^2\lambda_f^2)(1+3 k^2\lambda_f^2)}}
          -{\frac{6k^2{\dot{f}_R}{\lambda}_f\dot{\lambda}_f}{(1+f_R)(1+2 k^2\lambda_f^2)(1+3 k^2\lambda_f^2)}}.
          \label{bphi_c}
         \end{eqnarray}
 In these equations, $\dot{f}_R$, $\ddot{f}_R$, $\dot{\lambda}_f$, and $\ddot{\lambda}_f$ can be derived from background evolution without solving Eq.~(\ref{psi_2nd}) and Eq.~(\ref{phi_2nd}). In order to get a sense about the order of magnitude of these terms, we take the quasi-static approximation for $R$ and $\lambda_f$, and obtain $\dot{f}_R \sim f_{RR} \mathcal{H}^2R$, $\ddot{f}_R \sim (f_{RRR}+f_{RR}) \mathcal{H}^2R$, $\dot{\lambda}_f \sim \mathcal{H}{\lambda}_f$, and $\ddot{\lambda}_f \sim \mathcal{H}^2{\lambda}_f$, which we shall keep in mind are only approximately correct. In addition, for any viable $f(R)$ model, the function $f(R)$ can only grow steadily from an asymptotic constant with $f_R <0$ and $f_{RR}>0$, which implies that $|f_{RRR}|$ must be less than $f_{RR}$. Hence, when $f(R)$ effects kick in, $\dot{f}_R$ and $\ddot{f}_R$ are of the same order of magnitude as $\lambda^2_f$.

Based on the approximations above, it is reasonable to argue that the relationship between the exact but coupled second-order differential equations, Eq.~(\ref{EinEqijNew}) and Eq.~(\ref{EinEqii_alt}), and the two reduced decoupled differential equations, Eq.~(\ref{psi_2nd}) and Eq.~(\ref{phi_2nd}), depends on the competition between $k_X^2 \lambda_f^2$ and $1+k^2 \lambda_f^2$.
Indeed, the condition, $1+k^2 \lambda_f^2 \gg k_X^2 \lambda_f^2$, can be further broken down into $\lambda_f \ll 1/k_X$ or $k \gg k_X$, so either of which will be sufficient for the validity of Eq.~(\ref{psi_2nd}) and Eq.~(\ref{phi_2nd}).
Henceforth, we shall conclude that this transition from the exact equations to the reduced second-order ones will happen as long as either $f(R)$ effects become trivial ($\lambda_{f} \ll 1/k_X$) or the scale of the system is relatively small ($k \gg k_X$). We call $k \gg k_X$ \textit{the strong version of sub-Hubble limit} because this condition naturally contains the other two versions of sub-Hubble limit.
In the limit, $\lambda_{f} \ll 1/k_X$, all terms proportional to $\lambda_{f}$ or less in Eq.~(\ref{psi_2nd}) and Eq.~(\ref{phi_2nd}) are far less than $\mathcal{H}$, and $f(R)$ effects will become significant only when $k {\lambda_f}$ is nontrivial, that is, when $\lambda_f \gtrsim 1/k$. Similarly, in the limit $k \gg k_X$, $f(R)$ effects will dominate only when $k \gtrsim 1/\lambda_f$.

We should emphasize that although there are two separate conditions in our discussion for a given $f(R)$ model, the evolution of $\Psi$ and $\Phi$ might satisfy $\lambda_{f} \ll 1/k_X$ at the earlier time, and then satisfy $k \gg k_X$ only recently. If this is the case, and if time derivatives of $\Psi$ and $\Phi$ are negligible at later epoch, Eq.~(\ref{psi_2nd}) and Eq.~(\ref{phi_2nd}) will only require the weak version of the sub-Hubble limit, $k \gg \mathcal{H}_0$, for their validity, because during this period, $\mathcal{H}$ is compatible with $\mathcal{H}_0$.

So far we have overlooked what $k_X$ really represents by simply assuming that we can always find a $k$ that is large enough (or $\lambda_f$ that is small enough) to guarantee $k \gg k_X$ (or $\lambda_f \ll 1/k_X$). In fact, we may or may not find a case where the ratio between the time derivatives of potentials and potentials themselves, which we denote $\dot{X}/X$ and $\ddot{X}/X$, are so large that we will never find a reasonable $k$ or $\lambda_f$ for these approximations. To check this, we go back to analyze Eq.~(\ref{psi_2nd}) and Eq.~(\ref{phi_2nd}) themselves. If these equations yield highly unstable solutions (that is, $\Psi$ or $\Phi$ that increases/decreases significantly within a short period of time), our assumptions might be flaw; otherwise, they are fairly good. It turns out that for models that match $\Lambda$CDM at early time, a positive coefficient of the first derivatives in Eq.~(\ref{psi_2nd}) and Eq.~(\ref{phi_2nd}) will always keep the solutions stable, which naturally includes the cases when $\lambda_f \ll 1$. Indeed, when $\lambda_f \ll 1$, the potentials will hardly grow faster than $\ln{a}$; thus, the quasi-static approximation will never break down.
Even if $\dot{\abs{{f}_R}} \sim \lambda^2_f$ is compatible with $\mathcal{H}$, and makes the coefficient of the first derivatives in Eq.~(\ref{psi_2nd}) and Eq.~(\ref{phi_2nd}) negative (remember $\dot{f}_R<0$), while $k$ is large enough, $k^2 \gg \mathcal{H} \dot{X}/X$ or $\ddot{X}/X$ will hardly break down because the potentials must grow as fast as $e^{k^2\int{\mathcal{H}^{-1}}d\eta}$ to make $\dot{X}/X \sim k^2/\mathcal{H}$. Perhaps the only way to achieve this is when $\lambda_f \gg \mathcal{H}$, which clearly does not fit any viable $f(R)$ model.

\subsection{Quasi-Newtonian approximation for matter density contrast}\label{qs:m}

 Taking the strong version of sub-Hubble approximation, $k\gg k_X$, alone, we can drop all terms that are not proportional to $k^2$, so the equation of conservation of energy-momentum, Eq.~(\ref{eq:cons}), will be reduced to
        \begin{equation}
         {\ddot{\delta}}+\mathcal{H}{\dot{\delta}}+k^2 {\Psi}=0.
         \label{dE-M_appr}
         \end{equation}
Similarly, the Poisson equation, Eq.~(\ref{EinEq00Newf}), and the shear equation, Eq.~(\ref{EinEqijNew}), will lead to a simple relation between Newtonian potential and the matter density perturbation~\cite{zhang06,tsuji07},
         \begin{equation}
             \tilde{G} k^2 \Psi ={{8\pi G\rho_0}}\delta \;\;\;\;\;\;\; {\rm{and}} \;\;\;\;\;\;\;
              \tilde{G}\equiv-a (1+f_R){{2+6 k^2\lambda^2_f}\over{1+4k^2\lambda^2_f}},
             \label{delta_appr}
         \end{equation} where $\rho_0$ is a constant in the matter-dominated epoch. It shall be noticed that according to Eq.~(\ref{delta_appr}), $\Psi$ basically grows with a factor of $\delta/a$; thus, even if $\Psi$ behaves like a constant, time derivatives of $\delta$ could still be large.

          After putting Eq.~(\ref{delta_appr}) back into Eq.~(\ref{dE-M_appr}), we will obtain a second order differential equation for $\delta$, which often appears in the $f(R)$ literature because of the quasi-Newtonian approximation~\cite{bean07,cruz08,zhang06,tsuji07,tsuji08,tsujietal08}.
         However, as we have just shown,  Eq.(\ref{dE-M_appr}) and (\ref{delta_appr}) can be derived from the exact Poisson, shear and conservation equations, even if we take a more broad approximation, $k\gg k_X$, which includes the quasi-Newtonian approximation as well as the limit, $k \gg k_X \gtrsim \mathcal{H}$. In other words, as long as $k$ is large enough, Eq.(\ref{dE-M_appr}) and (\ref{delta_appr}) will always be correct, no matter if the quasi-static approximation breaks down or not.

         In order to compare  Eq.(\ref{dE-M_appr}) and (\ref{delta_appr}) with the reduced second-order differential equations in the last section, after some manipulation, we rewrite Eq.~(\ref{dE-M_appr}) in form of a second order differential equation of $\Psi$,
         \begin{equation}
         {\ddot{\Psi}}+\left(3\mathcal{H}+\tilde{A}_1-A_3\right){\dot{\Psi}}
           +\left(\dot{\mathcal{H}}+2\mathcal{H}^2+\tilde{A}_2+B_{\psi}\right){\Psi}=0,
         \label{psi_QS2}
         \end{equation}
         where $A_3$ is defined in Eq.~(\ref{a3_c}), $B_{\psi}$ in Eq.~(\ref{bpsi_c}), and
         \begin{eqnarray}
         \tilde{A}_1&=&{\frac{2\dot{f}_R}{1+f_R}},
         \\
          \tilde{A}_2&=&{\frac{3\mathcal{H}{\dot{f}_R}+{\ddot{f}_R}}{1+f_R}}
         -{\frac{(1+4k^2\lambda_f^2)a^2 4\pi G\rho}{(1+3 k^2\lambda_f^2)(1+f_R)}}
         -{\frac{4k^2{\dot{f}_R}{\lambda}_f\dot{\lambda}_f}{(1+f_R)(1+4 k^2\lambda_f^2)(1+3 k^2\lambda_f^2)}}.
         \end{eqnarray}
        A comparison between $\tilde{A}_1$, $\tilde{A}_2$, and $A_1$, $A_2$ shows that when $k{\lambda_f} \gg 1$ or $\dot{f}_R \ll1$, which also implies $\lambda_f \ll 1$, Eq.~(\ref{psi_2nd}) and Eq.~(\ref{psi_QS2}) will become identical.
        This is interesting because $\lambda_f \ll 1$ is included in the approximation, $\lambda_f \ll 1/k_X$, which is one of the two conditions that will reduce Eq.~(\ref{psi_2nd}). This means that when $\lambda_f$ is small enough, Eq.~(\ref{psi_QS2}), like Eq.~(\ref{psi_2nd}), will still be applicable even at near-Hubble or super-Hubble scales.
        This is in agreement with slow-rolling solutions in~\cite{nbf14}.
        On the contrary, when $\lambda_f$ is not so small, the condition $k{\lambda_f} \gg 1$ will be consistent with the strong version of sub-Hubble limit, $k\gg k_X$; thus, not surprisingly, Eq.~(\ref{psi_2nd}) and Eq.~(\ref{psi_QS2}) coincide when $k$ is large enough. In addition, just like Eq.~(\ref{psi_2nd}), when $\lambda_f \ll 1$, Eq.~(\ref{psi_QS2}) will always yield stable solutions, and quasi-static approximation will never break down; on the contrary, when $\lambda_f$ is large enough to have a negative coefficient in the first derivative term in Eq.~(\ref{psi_QS2}), but not so large to yield hyper-acceleratingly growing solutions, Eq.~(\ref{psi_QS2}) will always be a good approximation for $k\gg k_X$, regardless the validity of quasi-static approximation.

\subsection{Recovering Einstein-Hilbert Gravity}\label{GR_limit}

We have shown in Sec.~\ref{qs:p} that we can reduce the coupled pressure and shear equations into two independent second-order differential equations when $\lambda_f \ll 1/k_X$ or $k \gg k_X$. Here, we are going to show that we can even reduce these equations further into GR. Under the assumptions, $\lambda_{f} \ll 1/\mathcal{H}$ and $k \lambda_f \ll1$, all terms proportional to $\lambda_{f}$, including $k^2\lambda^2_{f}$, in Eq.~(\ref{psi_2nd}) and Eq.~(\ref{phi_2nd}) have only secondary effects compared to $\mathcal{H}$. Accordingly, both equations will simply recover the linear perturbation equation in standard gravity (see e.g.~\cite{Dodelson}),
         \begin{equation}
           {\ddot{\Psi}}+3\mathcal{H}{\dot{\Psi}}+(2\dot{\mathcal{H}}+\mathcal{H}^2){\Psi}=0,
           \label{psi_GR}
         \end{equation} where $\Psi=\Phi$ according to the shear equation, Eq.~(\ref{EinEqijNew}).

We can easily recognize that Eq.~(\ref{psi_GR}) is independent of scale, $k$. This differs from the case in $f(R)$ gravity, where the evolution of two potentials are scale-dependent. In addition to the scale-independence, Eq.~(\ref{psi_GR}) for standard gravity also stands out from its counterparts in $f(R)$ gravity in the following way.
According to Eq.~(\ref{psi_GR}), the impact of gravity on the potential only depends implicitly through $\mathcal{H}$. On the contrary, even in the sub-Hubble limit, the influence of $f(R)$ gravity is explicitly shown via $\lambda_f$ and its derivatives in Eq.~(\ref{psi_2nd}) and  Eq.~(\ref{phi_2nd}). In other words, even if two $f(R)$ models possess an identical cosmic background evolution, we can still in principle distinguish these two models because of their disparity on potentials. This is well known in the previous literature~\cite{song07}.

\subsection{Summary of different regimes of linear perturbations in $f(R)$ models}

\begin{figure}
\includegraphics[width=0.8\textwidth]{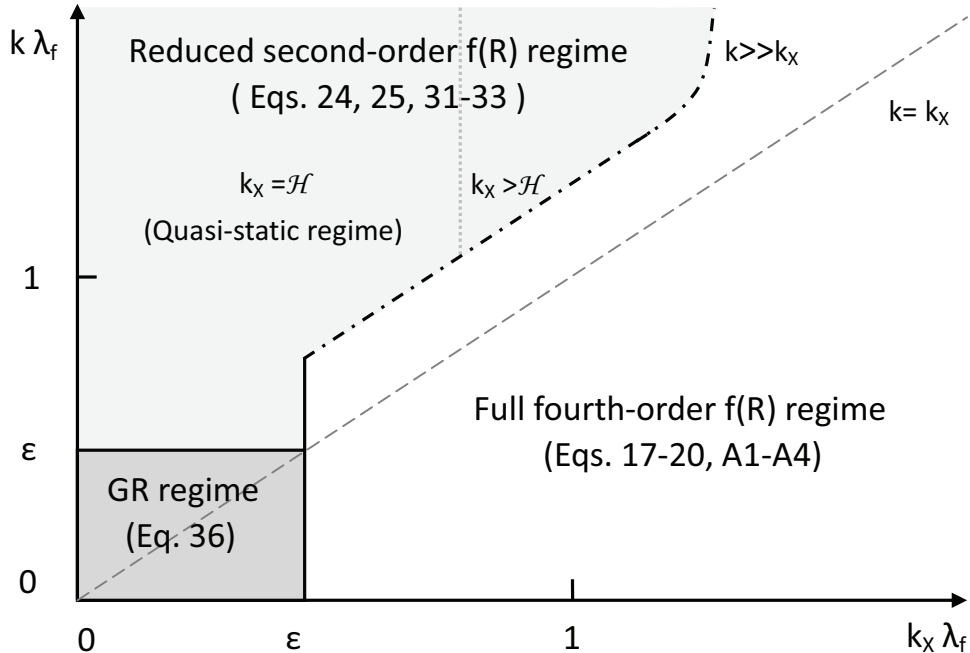}
\caption{Range of applicability of limits of the $f(R)$ equation. In this diagram, both axes increase with distorted scales, and we set $\varepsilon \ll1$. The dashed line represents $k=k_X$, and the dashed-dotted line, $k \gg k_X$. When  $k_X\lambda_f$ is not negligible, we enter the full fourth-order $f(R)$ regime, and only exact equations are applicable. In the limit $k_X\lambda_f \ll1$ or $k \gg k_X$, when $k\lambda_f$ is not negligible, the reduced second-order $f(R)$ regime is applicable. This regime could be further divided into the quasi-static ($k_X =\mathcal{H}$), and non-quasi-static ($k_X >\mathcal{H}$) subcategories, which is independent of $k$, and represented by the dotted line. In the limit ${\lambda_f \ll 1}$, when the regime of modified gravity vanishes, we return to the GR regime.}
\label{eq_compare}
\end{figure}

There is one scale length in $f(R)$ models, $\lambda_f$, which controls the scale on which the modified gravity effects contribute. Here, we are going to summarize how this scale length is connected to three different regimes of linear perturbations in $f(R)$ models, and show these connections in Figure~\ref{eq_compare}.

($1$) \textit{GR regime}. In the limit ${\lambda_f \ll 1}$, when $f(R)$ effects vanish, we return to the GR regime, Eq.~(\ref{psi_GR}). Indeed, ${\lambda_f \ll 1}$ is a necessary condition to guarantee both $k\lambda_f$ and $k_X\lambda_f$ to be trivial because the minimum value of $k_X$ is $\mathcal{H}$, and not negligible.

($2$) \textit{Reduced second-order $f(R)$ regime}. When $k\lambda_f$ is not trivial, but $k_X\lambda_f$, compared to $1$ or $k\lambda_f$, is still negligible, we will reach a regime where the exact equations will reduce to decoupled second-order equations, Eqs.~(\ref{psi_2nd}),(\ref{phi_2nd}),(\ref{dE-M_appr})-(\ref{psi_QS2}), which include the so called quasi-Newtonian equation for matter density contrast. Although in the previous literature the quasi-static approximation, $k_X=\mathcal{H}$, is assumed in order to derive the quasi-Newtonian equation for matter density contrast~\cite{cruz08,zhang06,tsuji07,tsuji08,tsujietal08,nbf14}, the quasi-static regime only constitutes part of the reduced second-order $f(R)$ regime. Indeed, as long as $k \gg k_X$, no matter the quasi-static approximation breaks down or not, the quasi-Newtonian equation for matter density contrast will always be valid. We would also like to emphasize that when $\lambda_f$ grows beyond a certain point, the quasi-static approximation will always break down ($k_X > \mathcal{H}$), and it will become more difficult to satisfy the condition, $k \gg k_X$.

($3$) \textit{Full fourth-order $f(R)$ regime}. When $k_X\lambda_f$ is not trivial compared to $1$ and $k\lambda_f$, it is impossible to further reduce the exact equations, Eqs.~(\ref{eq:cons})-(\ref{EinEqii_alt}),~(\ref{dotphi})-(\ref{doth}). The only possible way to turn these equations into decoupled ones is to recast them into fourth-order differential equations~\cite{cruz08,Chiu13}. We thus call this regime the full fourth-order $f(R)$ regime.

\section{Numerical Solutions of Linear Perturbations }\label{numerical}

In the previous sections, we have analytically shown that the pressure equation, Eq.~(\ref{EinEqii}), and the shear equation, Eq.~(\ref{EinEqijNew}), can be used to solve for Newtonian potential, and spatial curvature exactly. We also took these exact equations into various limits analytically. In this section, we are going to solve Eq.~(\ref{EinEqii}) and Eq.~(\ref{EinEqijNew}) numerically by turning these two coupled equations into four first-order differential equations, Eq.~(\ref{dotphi})-(\ref{doth}), and apply their numerical solutions  to check the potential disparity between the exact and quasi-Newtonian solutions at sub-Hubble scales, which is suggested by~\cite{cruz08}. When solving various equations for linear perturbations in $f(R)$ models, we have assumed that in the early epoch the evolution of both the Newtonian potential and the Newtonian spatial curvature follow that of standard gravity. Accordingly, without losing any generality, we normalize the potentials to $\Lambda$CDM at the very beginning, and choose $\Phi=\Psi=1$ and $\dot{\Phi}=\dot{\Psi}=0$ at $a=0.001$ as our initial conditions. In addition, given a specific $f(R)$ function, we use Eq.~(\ref{HubbleRatefR}) and Eq.~(\ref{BackGroundEv}) to obtain $\mathcal{H}$ and $R$ before solving for $\Psi$ or $\Phi$.

\subsection{Comparisons between solutions of exact equations and various limits }

To solve equations of linear perturbations, we still need to specify $f(R)$ functions, and know the Hubble parameter beforehand. For this purpose, we apply two specific $f(R)$ models, ${f_a(R)=-6 \Omega_{\Lambda}+(R/R_0)^{-1.5}}$
and ${f_b(R)=-5 \Omega_{\Lambda}+10(R/R_0)^{-1.5}}$, where $R_0\equiv c^{-2}\mathcal{H}_0^2$, $c=1$, and $\mathcal{H}_0=1$. We use these two models as exemplars because they both yield reasonable background evolution, but at the same time have significant distinguishable linear perturbations compared to $\Lambda$CDM.
Moreover, since both $f_{R}$ and $f_{RR}$ of $f_b(R)$ are ten times larger than that of $f_a(R)$, a comparison of linear perturbations between these two models are going to provide us with some hint about how $f_{R}$ and $f_{RR}$ will affect linear perturbations in $f(R)$ gravity.

\begin{figure}
\includegraphics[width=0.8\textwidth]{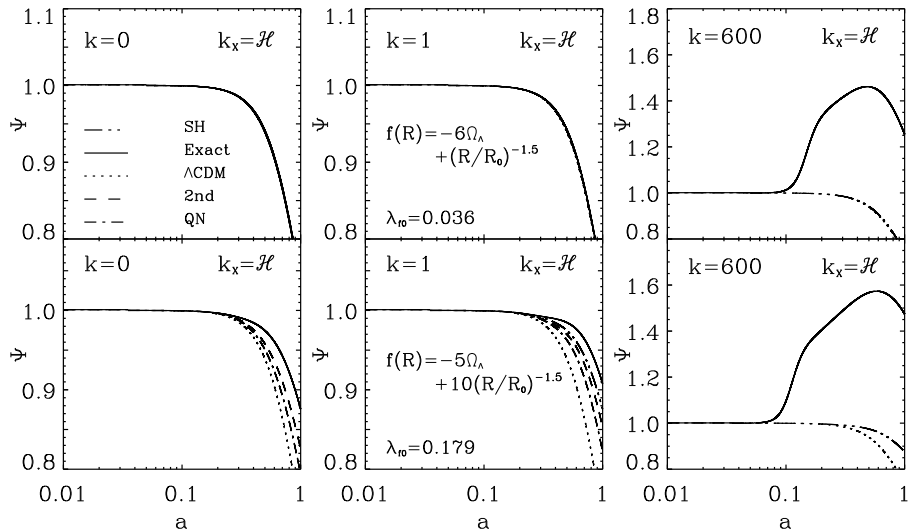}
\caption{\label{compare_gamma_QS}  Example of comparisons of the evolutions of $\Psi$. The models, ${f_a(R)=-6 \Omega_{\Lambda}+(R/R_0)^{-1.5}}$, and ${f_b(R)=-5 \Omega_{\Lambda}+10(R/R_0)^{-1.5}}$ are used to compare the evolutions of $\Psi$ solved from different schemes at three scales, $k=0$, $1$, and 600 (in units, $c^{-1}\mathcal{H}_0$), where $R_0\equiv c^{-2}\mathcal{H}_0^2$, and we set $c=1$ and $\mathcal{H}_0=1$.
Here, $\lambda_{f0}$ is defined as the scale length, $\lambda_f$, at $a=1$, and $k_X=\mathcal{H}$ for all $k$ in the both models. These schemes include the exact equations (exact), Eq.~(\ref{dotphi})-(\ref{doth}), the reduced second-order differential equation (2nd), Eq.~(\ref{psi_2nd}), $\Lambda$CDM, Eq.~(\ref{psi_GR}), the quasi-Newtonian equation (QN), Eq.~(\ref{psi_QS2}), and the super-Hubble equations (SH) in Appendix~\ref{App_B}, Eq.~(\ref{phi_psi_hu})-(\ref{B_lenth}).
For the first model, the solutions of exact, 2nd, and QN agree with each other at all scales. Although these solutions also agree with $\Lambda$CDM and SH at $k=0$ and $1$, they deviate significantly at $k=600$. Also, for this model, $\Lambda$CDM agrees with SH at all scales. For the second model, the solutions of exact, 2nd, and QN differ at $k=0$ and $1$, but still agree at $k=600$. Unlike the first model, $\Lambda$CDM differs from all other solutions, even including SH. On the contrary, the SH line is under the exact line at $k=0$. The differences between these two models are caused by the fact that $f_R$ and $f_{RR}$ of the second model are $10$ times larger than the first one.}
\label{compare_psi2}
\end{figure}

In Fig.~\ref{compare_psi2}, we compare the Newtonian potentials solved for the models $f_a(R)$ and $f_b(R)$ from four exact first-order equations, Eq.~(\ref{dotphi})-(\ref{doth}), from the reduced second-order equations,  Eq.~(\ref{psi_2nd}), and from the quasi-Newtonian second-order differential equations, Eq.~(\ref{psi_QS2}). We plot the solutions at three scales, $k=0$, $1$, and $600$ (remember $k$ is in units of $c^{-1}\mathcal{H}_0$). We can see that all the Newtonian potentials solved from these three different ways concur at $k=600$, where $k \gg \mathcal{H}$, and at the early Universe, where $\lambda_f \ll1/\mathcal{H}$. This concurrence continues at the super-Hubble scale, $k=0$, as well as the near-Hubble scale, $k=1$, for the model $f_a(R)$, but breaks down at the same scales for the model $f_b(R)$. As we have shown at the end of Section~\ref{limit}, this difference can be explained by the fact that $\lambda_f \ll 1/\mathcal{H}$ is still valid in the model $f_a(R)$ at the recent epoch ($\lambda_f=0.036$ at $a=1$), but not in the model $f_b{(R)}$ ($\lambda_f=0.179$ at $a=1$).

In Section~\ref{limit}, we also have shown that under the limit $\lambda_f \ll1$, linear perturbation in $f(R)$ gravity will recover standard gravity. This, however, will break down if the perturbations enter the regime, $k^{-1} \ll \lambda_f$.
This conclusion is again supported by Fig.~\ref{compare_psi2}. In this figure, the Newtonian solutions solved from the exact equation, the reduced second-order equations, and the quasi-Newtonian second-order differential equations all coincide with $\Lambda$CDM for the model $f_a(R)$ at $k=0$, and $1$. This consistency then breaks down at $k=600$ because the effects of $f(R)$ gravity caused by $k \lambda_f$ are no longer negligible at this scale. On the contrary, for the model, $f_b{(R)}$, the effects of $f(R)$ gravity are nontrivial at all scales because this model possesses the larger value of $\lambda_f$.

In addition to the approximated equations in the sub-Hubble limit as we have discussed in this paper,~\cite{song07} offers us another approximated equations, Eq.~(\ref{phi_psi_hu})-(\ref{B_lenth}), for the opposite end of the scale, $k=0$. In Fig.~\ref{compare_psi2}, we also compare the solutions of this approximated equations for $k=0$ with our exact solutions solved from Eq.~(\ref{dotphi})-(\ref{doth}). The comparison shows that the Newtonian potentials solved from these two different sets of equations do match at $k=0$, but not necessarily in the sub-Hubble limit. All these details prove the consistency between our exact equations and Eq.~(\ref{phi_psi_hu})-(\ref{B_lenth}) at $k=0$.

\begin{figure}
\includegraphics[width=0.8\textwidth]{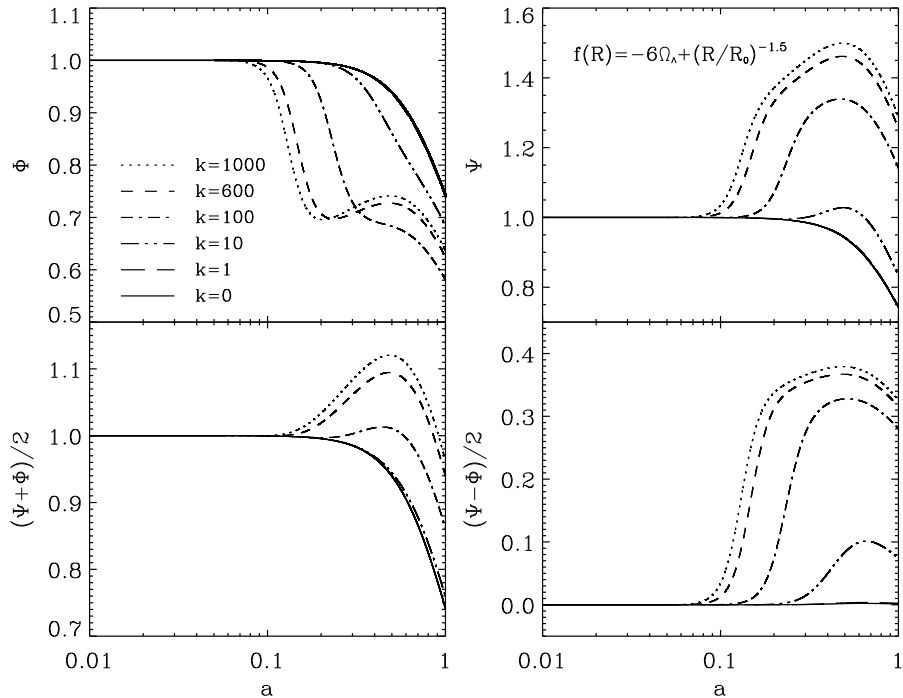}
\caption{Solutions for exact linear perturbations. The exact solutions of $\Psi$ and $\Phi$ are solved from four first-order differential equations, Eq.~(\ref{dotphi})-(\ref{doth}), for the model, ${f_a(R)=-6 \Omega_{\Lambda}+(R/R_0)^{-1.5}}$, between $k=0$ and $1000$ (in the units, $c^{-1}\mathcal{H}_0$); $R_0\equiv c^{-2}\mathcal{H}_0^2$, where we set $c=1$, and $\mathcal{H}_0=1$.
For this $f(R)$ model, $\Psi$ and $\Phi$ at $k=0$ are indistinguishable from their counterparts at $k=1$, and $\Lambda$CDM.
The combination of the two potentials, $(\Psi+\Phi)/2$, are shown in the left-lower panel. In the right-lower panel, we also plot $(\Psi-\Phi)/2$; their counterparts in $\Lambda$CDM are always zero. Because larger $k$ will become compatible with the Compton length, $\lambda_f$ earlier than smaller $k$, $f(R)$ effects will start earlier at sub-Hubble scales. Because effects of $f(R)$ are proportional to $k^2 \lambda_f^2$ at near or sub-Hubble scales, potentials at smaller scales will deviate from $\Lambda$CDM more dramatically than their counterparts at Hubble or super-Hubble scales, where the effects of $f(R)$ gravity are only proportional to $\lambda_f^2$.}
\label{compare_k_v2}
\end{figure}

\subsection{Evolutions of $\Psi$  and $\Phi$ in $f(R)$ gravity}

In the last section, we compare the solutions of the exact equations with various limits, including $\lambda_f \ll 1$, $k \gg 1$, $k\lambda_f \gg 1$, and $k=0$. The approximate solutions agreed with the exact solutions in the appropriate limit. In Fig.~\ref{compare_k_v2}, we show that the evolutions of $\Phi$, $\Psi$, $(\Psi+\Phi)/2$ and $(\Psi-\Phi)/2$ solved from the exact equations Eq.~(\ref{dotphi})-(\ref{doth}) at several scales for the model $f_a(R)$. Unlike the linear perturbation in $\Lambda$CDM, where evolutions of potential are independent of $k$, both the Newtonian potential and curvature potential at sub-Hubble scales ($k\gg1$) deviate significantly from their counterparts at super-Hubble scales.
This can be explained by the fact that the effects of $f(R)$ gravity are related to $k\lambda_f$ at sub-Hubble scales, but are only related to $\lambda_f$ at super-Hubble scales.

\begin{figure}
\includegraphics[width=0.6\textwidth]{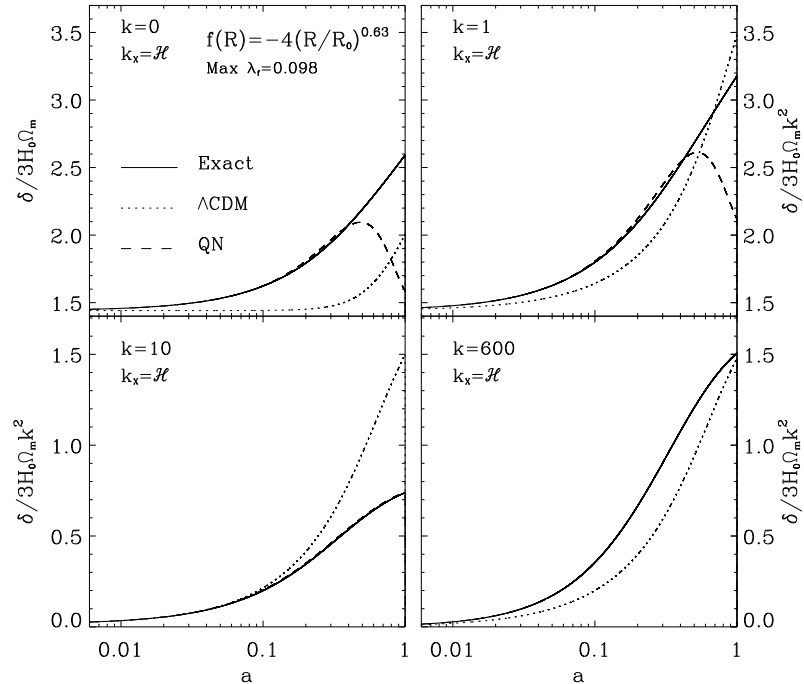}
\caption{\label{compare_delta} Comparisons of $\delta$ for ${f(R)=-4(R/R_0)^{0.63}}$, where $R_0\equiv c^{-2}\mathcal{H}_0^2$, and we set $c=1$, and $\mathcal{H}_0=1$. The perturbations of the matter density in ${f(R)=-4(R/R_0)^{0.63}}$ are solved from the exact equations (Eq.~(\ref{EinEq00Newf}) and Eq.~(\ref{dotphi})-(\ref{doth}); exact), the quasi-Newtonian equation for $\delta$ (combination of Eq.~(\ref{dE-M_appr}) and Eq.~(\ref{delta_appr}) ; QN), and $\Lambda$CDM for $k=0$, $1$, $10$, and, $600$ (in units, $\mathcal{H}_0/c$). Although quasi-Newtonian solutions (QN) do not fit the exact solutions (exact) at super-Hubble scale, $k=0$, or around Hubble scales, $k=1$, there is no obvious disparity between two solutions at near, $k=10$, nor sub-Hubble scales, $k=600$ $\mathcal{H}_0/c$. In this model, $k_X=\mathcal{H}$ for all $k$, and the quasi-static assumption has never broken down.}
\end{figure}

In Fig.~\ref{compare_k_v2} we also show that there exists significant differences between the Newtonian potential and curvature potential.
At sub-Hubble scales, the Newtonian potentials will mostly grow up from a constant, reach a peak, and then decline almost at the same rate for all scales. The curvature potentials at sub-Hubble scales will decline faster than their counterparts at super-Hubble scales, oscillate, and eventually, like the Newtonian potentials, decline almost at the same rate for all scales.
Because for the model $f_a(R)$, the exact solutions are identical to the solutions of the reduced second-order differential equations, Eq.~(\ref{psi_2nd}) and Eq.~(\ref{phi_2nd}), it is possible to understand why the Newtonian and curvature potentials evolve in their particular ways by analyzing  Eq.~(\ref{psi_2nd}) and Eq.~(\ref{phi_2nd}) themselves. In this way, we found that the coefficients of the $\dot{\Psi}$ and $\dot{\Phi}$ terms are always positive
for the model $f_a(R)$, so these terms only contribute to damp the evolutions of both potentials. The negative and positive parts of the coefficients of $\Psi$ and $\Phi$ contribute to boost and suppress the evolutions of $\Psi$ and $\Phi$ at sub-Hubble scales.

\subsection{Issues around quasi-static assumption}

As we have mentioned before,~\cite{cruz08} has found a fourth-order differential equation to solve for matter density perturbation exactly. By taking the sub-Hubble limit, they reduce their fourth-order differential equation into a second-order one, which they claim is not necessarily identical Eq.~(\ref{dE-M_appr}) combined with Eq.~(\ref{delta_appr}) even at sub-Hubble scales. They contribute this disparity to the over-aggressive quasi-static approximation behind the quasi-Newtonian equations Eq.~(\ref{dE-M_appr}) and Eq.~(\ref{delta_appr}).

In order to check their claims, we solve for $\Psi$ and $\Phi$ from the exact equations, Eq.~(\ref{dotphi})-(\ref{doth}) for the model $f_c(R)=-4(R/R_0)^{0.63}$, which is used arbitrarily by~\cite{cruz08} to show the disparity between the quasi-Newtonian and the exact solutions at $k=600$. We then apply these solutions back to Poisson equation, Eq.~(\ref{EinEq00Newf}), to obtain matter density contrast, $\delta$. We are confident in our solutions because all of them pass the test of consistency by putting our solutions back into the shear and pressure equations, Eq.~(\ref{EinEqii}) and Eq.~(\ref{EinEqijNew}).
We compare this $\delta$ with the matter density contrast solved from Eq.~(\ref{dE-M_appr}) combined with Eq.~(\ref{delta_appr}), and plot the results in Fig.~\ref{compare_delta} for several scales.

Fig.~\ref{compare_delta} shows that although there exits significant disparity between the exact solutions and the quasi-Newtonian ones for $k<1$, this disparity vanishes completely for $k>10$, which, of course, includes $k=600$. This result conforms with our analysis in Section~\ref{limit}, where we show that the difference between the exact solutions and the approximated solutions of Eq.~(\ref{psi_2nd}), Eq.~(\ref{phi_2nd}), and Eq.~(\ref{psi_QS2}) are solely determined by the competitive terms of $\lambda_f$ and $k \lambda_f$ in the shear equation, Eq.~(\ref{EinEqijNew}). When $k$ becomes as large as $10$ and $\lambda_f$ is not really so large (the maximum value of $\lambda_f$ for the model $f_c(R)$ is $0.098$ between the CMB is formed and today), the reduced second-order differential equations (or quasi-Newtonian equations) will be fairly good approximations.

\section{Discussion and conclusion}\label{conclusion}

In this paper, we provide a new insight into how to solve for linearized cosmological perturbations in $f(R)$ models correctly and efficiently.
Although in principle, four among two conservation equations and four modified Einstein equations of linearized cosmological perturbations are sufficient to solve for the evolutions of $\Phi$, $\Psi$, $\theta$ and $\delta$ in $f(R)$ gravity, in practice it is hard to solve these coupled equations simultaneously. Many studies circumvent this problem by taking insightful assumptions for some specific circumstance, such as quasi-static or sub-Hubble approximations~\cite{zhang06,bean07,tsuji07,amen08,tsuji08,tsujietal08}. In this paper, we analytically prove that without losing any information, the pressure equation, Eq.~(\ref{EinEqii}), and the shear equation, Eq.~(\ref{EinEqijNew}), alone are sufficient to solve for $\Phi$ and $\Psi$, which although is obvious in GR, has never been shown explicitly in the $f(R)$ literature. After obtaining evolutions of $\Psi$ and $\Phi$, we can easily derive $\delta$ and $\theta$ from the Poisson equation, Eq.~(\ref{EinEq00Newf}), and the momentum equation, Eq.~(\ref{EinEq0if}).

One of the advantages to apply the pressure equation, Eq.~(\ref{EinEqii}), and the shear equation, Eq.~(\ref{EinEqijNew}), rather than the equations used in~\cite{song07,bean07,libarrow07,pog08,cruz08,ceo10}, is that we can easily understand how Eq.~(\ref{EinEqii}) and Eq.~(\ref{EinEqijNew}) will reduce into the decoupled second-order differential equations, Eq.~(\ref{psi_2nd}) and Eq.~(\ref{phi_2nd}),  which themselves can be applied for the consistent analysis of the quasi-static approximation. In our reductions, we naturally find a Compton length, $\lambda_f$, to characterize the effects of $f(R)$ gravity in linear perturbations. We conclude how the exact equations, Eq.~(\ref{EinEqii})-(\ref{EinEqijNew}), fit into the reduced second-order differential equations Eq.~(\ref{psi_2nd})-(\ref{phi_2nd}), quasi-Newtonian equations, Eq.~(\ref{delta_appr})-(\ref{psi_QS2}), and Einstein gravity, Eq.~(\ref{psi_GR}), solely depends on the effects of the $\lambda_f$ and $k \lambda_f$ terms in the shear equation, Eq.~(\ref{EinEqijNew}). We reach the same conclusion of~\cite{song07,hs07} from a different approach.

We also find that $\lambda_f$ plays an important role in the quasi-static approximation: when $\lambda_f$ is small, which is applicable for most if not all observational viable $f(R)$ models~\cite{hs07,Schmidt09,LSSH12,planck15,brax08,sch09,ferr11,jain12,lombetal12}, quasi-static approximation will always be valid. Our finding agrees well with~\cite{bhb15}, where the quasi-static approximation has been proved valid for $\abs{f_R}=10^{-4}-10^{-6}$. Because small $\lambda_f$ will generically result in $\Lambda$CDM-like background evolutions, our conclusion is also consistent with~\cite{cruz08,nbf14}, which find that $\Lambda$CDM-like backgrounds will guarantee the applicability of quasi-static approximation. However, we would like to emphasize that although a small Compton length will guarantee a proximity of background evolutions to $\Lambda$CDM, it is not always correct the other way around. As shown in~\cite{song07}, different $f(R)$ models might lead to the same background evolution that is identical to $\Lambda$CDM. Our analysis shows that it is $\lambda_f$ rather than a proximity of background evolutions to $\Lambda$CDM that will guarantee the quasi-static approximation.
This finding might even have a deep connection to the conclusion of~\cite{sb15}, which shows that the quasi-static behaviors are related to the sound horizon of dark energy or modified gravity; that is, the nature of models themselves.
We would like to explore this plausible connection in the future.

In addition to quasi-static approximation, we also discuss the role of sub-Hubble limit in taking approximations of the exact equations, Eq.~(\ref{EinEqii})-(\ref{EinEqijNew}).
Our analysis also shows that whether or not quasi-static approximation beaks down, as long as $\lambda_f$ is not extremely large we can always find a $k$ that is large enough to guarantee the reduced second-order differential equations, Eq.~(\ref{psi_2nd})-(\ref{phi_2nd}), and quasi-Newtonuan equations, Eq.~(\ref{delta_appr})-(\ref{psi_QS2}). Indeed, even for the model that~\cite{cruz08} found problematic in their quasi-Newtonian solutions at $k=600$, we cannot find any obvious disparity between the exact and quasi-Newtonian solutions up to $k=10$.
Our conclusion contradicts~\cite{cruz08}, but is consistent with~\cite{hpst12}.

Generally speaking, the pressure equation, Eq.~(\ref{EinEqii}), and the shear equation, Eq.~(\ref{EinEqijNew}) are only based on the assumption of a flat RW metric and a matter-dominated era. Therefore combined with Poisson equation, these two equations shall be legitimate to be applied to investigate oscillating solutions of matter density perturbations. Since our analysis casts doubts in some conclusions in~\cite{cruz08}, it will be interesting to re-analyze these oscillating solutions from our approach, and compare them with the results in~\cite{mat13}.
We leave the exploration of this study to the near future.

\begin{acknowledgments}
We would like to thank John Peacock, Alan Heavens, and Fergus Simpson for their helpful discussions.
We also like to thank Pedro Ferreira for his helpful comments on this work.
MCC is partly supported by SUPA(UK). This work is also partly supported by the Chinese National Nature Science foundation Nos. 11433003 \& 11333001, Shanghai Science Foundations No. 13JC1404400.
\end{acknowledgments}

\appendix

\section{Exact Decoupled Equations for  Linear Perturbations in $f(R)$ Gravity}\label{4thODE}

We have seen the pressure equation, Eq.~(\ref{EinEqii}), and the shear equation, Eq.~(\ref{EinEqijNew}), are sufficient to solve for $\Phi$ and $\Psi$ exactly. Practically speaking, in order to solve for these two potentials, we might apply numerical method by turning Eq.~(\ref{EinEqii}) and Eq.~(\ref{EinEqijNew}) into four coupled first-order differential equations:
         \begin{eqnarray}
            \dot{\Psi}&=&\mathfrak{g},
            \label{dotphi}
            \\
            \dot{\Phi}&=&\mathfrak{h},  \\
            \dot{\mathfrak{g}}&=& \mathfrak{G}(\mathfrak{h},\mathfrak{g},\Psi,\Phi), \\
            \dot{\mathfrak{h}}&=& \mathfrak{H}(\mathfrak{h},\mathfrak{g},\Psi,\Phi),
            \label{doth}
         \end{eqnarray} where,
         \begin{eqnarray}
            \mathfrak{G}&=&\Big({a^2(1+f_R)\over 6 f_{RR}}+{2\over3}k^2-\dot{\mathcal{H} }-2 \mathcal{H}^2+{{\mathcal{H} \dot{f_R}+\ddot{f_R}} \over {1+f_R}}\Big)\Phi\nonumber \\
            &&-\Big({a^2(1+f_R)\over 6 f_{RR}}+{1\over3}k^2-\dot{\mathcal{H}}+{{5\mathcal{H} \dot{f_R}+2\ddot{f_R}-a^2 8\pi G \rho} \over {1+f_R}}\Big)\Psi -\Big(2\mathcal{H}+3{{\dot{f_R}}\over{1+f_R}}\Big)\mathfrak{g},
             \label{vG}
            \\
                \mathfrak{ H}&=&-\Big({a^2(1+f_R)\over 6 f_{RR}}+{2\over3}k^2\Big)\Phi+\Big({a^2(1+f_R)\over 6 f_{RR}}+{1\over3}k^2-2(\dot{\mathcal{H}}+\mathcal{H}^2)\Big)\Psi-3\mathcal{H}\mathfrak{h}
                -\mathcal{H}\mathfrak{g}.
            \label{vH}
         \end{eqnarray}
Although in practical Eq.~(\ref{dotphi})-Eq.~(\ref{vH}) are applied to solve for the Newtonian potential, $\Psi$, and the curvature potential, $\Phi$, numerically, the pressure equation, Eq.~(\ref{EinEqii}), and the shear equation, Eq.~(\ref{EinEqijNew}), themselves will provide us with a more intuitive sense about how the effects of $f(R)$ gravity will contribute to the linearized cosmological perturbations.

\section{The Super-Hubble limit}\label{App_B}
$\;\;\;\;$
In their paper about the large scale structure of $f(R)$ models,~\cite{song07} offers us
a simple relation between $\Psi$ and $\Phi$ at $k=0$:
         \begin{equation}
         \Psi={\Phi+a B \Phi'\over B-1}={\mathcal{H}\Phi+ B \dot{\Phi}\over \mathcal{H}(B-1)}.
                   \label{phi_psi_hu}
         \end{equation}

They also provide us an equation that can exactly describe the linear perturbation at the super-Hubble scale, $k=0$. In our notation this equation appears as
        \begin{eqnarray}
         {\ddot{\Phi}}&+&\Big(\mathcal{H}-{\dot{\mathcal{H}}\over\mathcal{H}}+{2\over a}
               +{{\ddot{\mathcal{H}}\mathcal{H}-2\dot{\mathcal{H}}\mathcal{H}^2-\dot{\mathcal{H}}^2}
                                          \over{\dot{\mathcal{H}}\mathcal{H}^2-2\mathcal{H}^4}}
               +{\dot{B}\over{a{\mathcal{H}}(1-B)}}-{B\over a}+{B\dot{\mathcal{H}}\over a\mathcal{H}^2} \Big){\dot{\Phi}} \nonumber \\
         &+&\Big({- \mathcal{H}\over a}+{\dot{\mathcal{H}}\over a \mathcal{H}}
               +{{\ddot{\mathcal{H}}\mathcal{H}-2\dot{\mathcal{H}}\mathcal{H}^2-\dot{\mathcal{H}}^2}
                                          \over{\dot{\mathcal{H}}\mathcal{H}-2\mathcal{H}^3}}
               +{\dot{B}\over a({1-B})}
         \Big)\Phi=0.
         \label{phi_super}
         \end{eqnarray} where
         \begin{equation}
         B={f_{RR}\over{1+f_R}}R'{a\mathcal{H}\over {a\mathcal{H}'-\mathcal{H}}}={f_{RR}\over{1+f_R}}\dot{R}{\mathcal{H}\over {\dot{\mathcal{H}}-\mathcal{H}^2}}
         \label{B_lenth}
         \end{equation} is a specific parameter used by~\cite{song07}, and $B^{1/2}$ is called Compton length by~\cite{hs07}.
 Analytically, there seems no obvious way to prove that the pressure and shear equations can be reduced into Eq.~(\ref{phi_psi_hu}) and Eq.~(\ref{phi_super}) at $k=0$. However, our numerical solutions in Sec.~\ref{numerical} show that the Newtonian potentials from both sides do concur at this very super-Hubble scale.

{}

\end{document}